\documentclass[10pt]{iopart}

\usepackage[colorlinks=true,citecolor=blue,menucolor=blue,linkcolor=blue,urlcolor=blue]{hyperref}

\usepackage{amsfonts}
\usepackage{amssymb}
\usepackage{longtable}
\usepackage{graphicx}
\usepackage[tabke]{xcolor}
\usepackage{multirow}
\usepackage{color}

\renewcommand{\Re}{\mathop{\mathrm{Re}}}
\renewcommand{\Im}{\mathop{\mathrm{Im}}}

\begin{document}

\title{Optical properties of graphene quantum dots: the role of chiral symmetry}

  \author{Denis M Basko$^1$, Ivan Duchemin$^2$, Xavier Blase$^3$}
  \address{$^1$ Universit{\'e} Grenoble Alpes and CNRS, LPMMC, 
  38042 Grenoble, France} 
  \address{$^2$ Universit{\'e} Grenoble Alpes and CEA, IRIG-MEM-L$\_$Sim, 38000 Grenoble, France}
  \address{$^3$ Universit{\'e} Grenoble Alpes and CNRS, Institut N\'eel, 38042 Grenoble, France}


\begin{abstract}
We analyse the electronic and optical properties of graphene quantum dots (GQD)  using accurate \textit{ab initio} many-body $GW$ and  Bethe-Salpeter calculations. We show that most pristine GQD, including structures with irregular shapes, are characterized by  dark low energy singlet excitations that quench fluorescence. We rationalizqe this property by exploiting the  chiral symmetry of the low energy electronic states in graphene. Edge  \textit{sp}$^3$ functionalization is shown to efficiently brighten these low lying  excitations by  distorting  the \textit{sp}$^2$ backbone planar symmetry. Such findings   reveal  an original indirect scenario for the influence of functionalization on the photoluminescence properties. 
\end{abstract}

\submitto{\TDM}
\maketitle
\ioptwocol

\section{Introduction}
The fluorescence properties of graphene quantum dots (GQDs), namely small-size monolayer or multilayer graphene flakes \cite{Pan10,Li10,Li11,Zhu11,Shen12,Peng12,Tang12,Jin13,Li13,Wan14,Lim15,Ding16},  are attracting significant interest for potential applications  in optoelectronic \cite{Lim15} including bioimaging~\cite{Zhu11,Shen12,Peng12,Li13,Wan14,Ding16},  photovoltaic~\cite{Li11,Wan14}, sensing~\cite{Shen12,Li13} or photocatalytic \cite{Li10,Shen12,Wan14} devices. 
The nature of the emitting states, at the core of these properties, is hindered by the large variety of top-down and bottom-up available synthesis techniques. The possible candidates   may range from intrinsic $\pi-\pi^*$ transitions, in a confined $sp^2$ system,  to edge states, including e.~g. oxygen-rich functional groups or carbene-like zigzag sites.
As a result, the main factors influencing the emission wavelength are still much debated.  


Pristine GQD properties have been explored at the density functional theory (DFT) and time-dependent TD-DFT levels clearly emphasizing the opening of the photoemission and optical gaps by quantum confinement with decreasing GQD size~\cite{Eda10,Sk14}. Further work demonstrated  that  functionalization~\cite{Cocchi2012, Jin13,Sk14,Gee16,Che18,Wan18}   and/or doping \cite{Sk14,Niu16,Noo16,Fen18,Kadian_2019} can significantly affect the electronic and optical properties of GQDs. While such studies shed some light on the large variety of photoluminescence properties that can be observed experimentally given the chosen synthesis route and edge treatment, some intriguing properties of pristine GQDs were reported \cite{Hei08, Sch11, Zhao2014, Ozfidan2014, Li15, Pohle2018}. In particular, the lowest excitations were found to have very small optical transition dipoles. This can manifest itself in a large Stokes shift between absorption and luminescence peaks, or, if efficient non-radiative decay channels are present, in quenching of the photoluminescence. These properties were associated with the high geometrical symmetry of the considered ideal flakes~\cite{Sch11, Ozfidan2014, Pohle2018}.  

\begin{figure}
 \includegraphics[width=8cm]{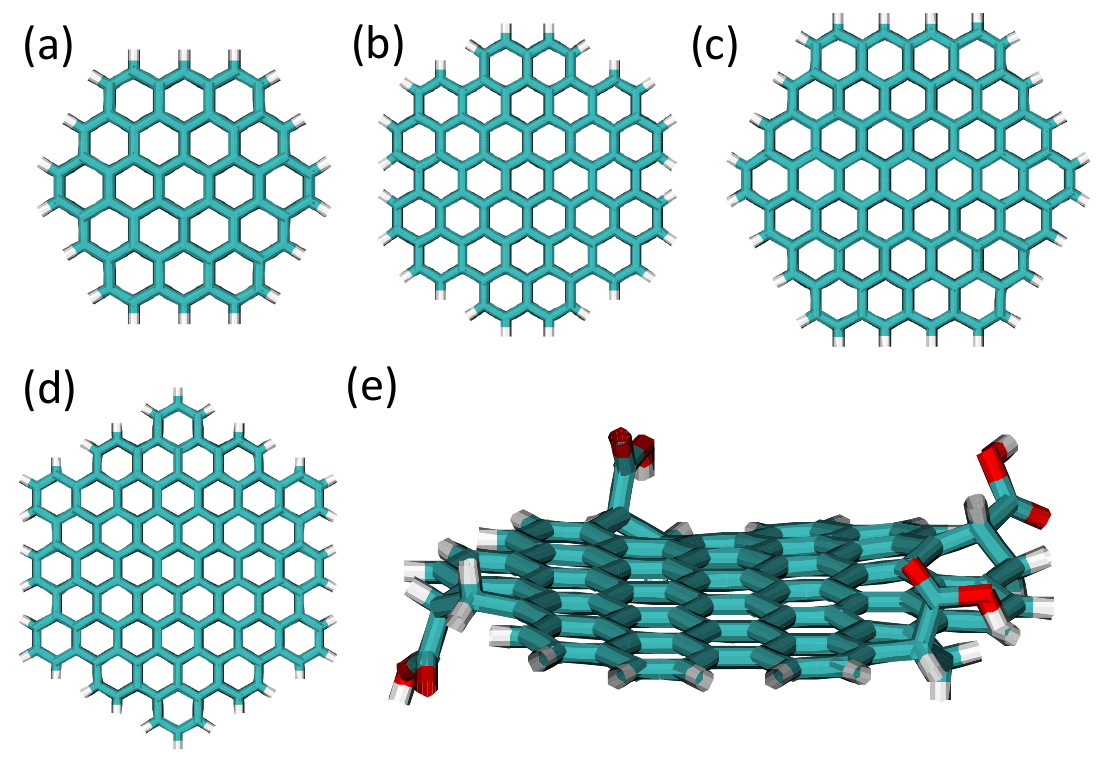} 
  \caption{High-symmetry GQDs with (a)  72, (b)  108, (c) 120 and (d) 144 total number of atoms.   Structure (e) is the 108-atoms GQD  with four -COOH groups functionalizing  H-passivated C-atom resulting in a local $\textit{sp}^3$ configuration. Relative sizes are not respected. }
\label{fig1}
\end{figure}    
In the present study, we show that the presence of low lying dark excitations in pristine GQDs is a general property rooted in the hexagonal symmetry of the underlying graphene lattice and the electron-hole chiral symmetry. Moreover, this property is preserved also for structures deviating significantly from high symmetry shapes. These conclusions are confirmed by \textit{ab initio} many-body Green's function calculations performed on realistic GQDs.  We argue that the chiral symmetry imposes a certain hierarchy of energy scales which (i)~persists even when the spatial symmetries are lifted and (ii)~results in the lowest singlet excitation being dark. We show that $sp^3$ functionalization of the kind observed experimentally very efficiently brightens the low lying dark states by breaking the planarity of the GQDs. As such, edge functional groups strongly impact indeed the GQDs photoluminescence even though not contributing directly to the emission process.

\begin{figure*}
 \includegraphics[width=\textwidth]{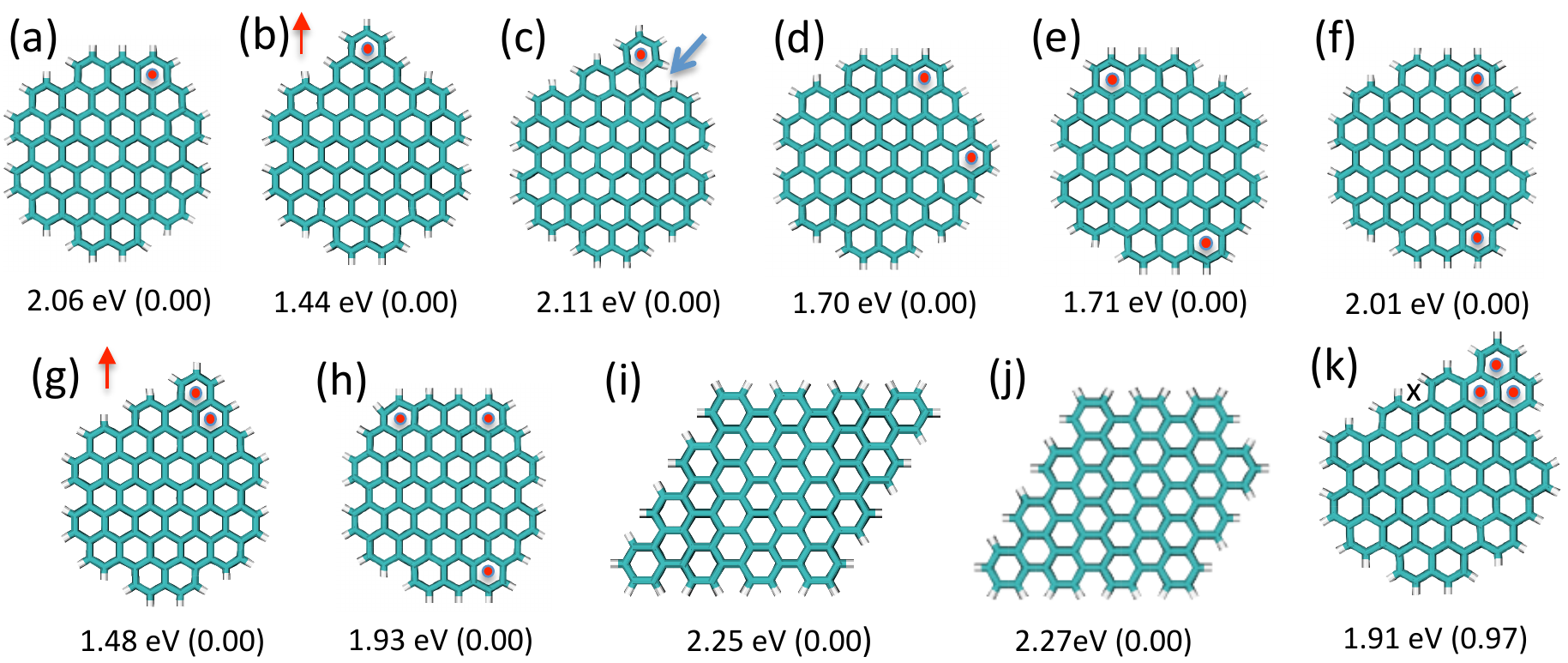} 
  \caption{Structures obtained from the 108-atoms GQD of Fig.~\ref{fig1}(b) by adding one~(a-c), two~(d-g) or three~(h) rings (red dots), or by removing one ring (cross). Structures (b,g) are open-shell (indicated by red arrows) while in (c) the blue arrow indicates a steric conflict. The associated lowest singlet ($S_1$) TD-PBE0 excitation energy and oscillator strength (in parenthesis) are indicated below each structure.  Information about higher excitations is provided in  \ref{BSEeigenstates}. }
    \label{fig2}
\end{figure*}


\section{Methodology}
The electronic and optical properties of pristine and functionalized GQDs are studied using the \textit{ab initio} many-body $GW$ \cite{Hedin65} and Bethe-Salpeter equation (BSE) formalisms~\cite{Sal51,Hanke79}.
The BSE approach was shown in particular to provide a balanced and accurate description of both Frenkel and charge-transfer optical singlet excitations in molecular systems~\cite{Bla18}.
Our calculations are performed with the {\sc{Fiesta}} package \cite{Jac15,Duc18} at the 6-311Gd eigenvalue-self-consistent  ev$GW$@PBE0 level, 
where the corrected electronic energy levels are re-injected self-consistently in the construction of $G$ and $W$  for sake of accuracy~\cite{Kap16,Ran16}.
 Our calculations are performed using the Coulomb-fitting resolution of the identity (RI-V) \cite{Ren12,Duc17} together with the auxiliary Weigend Coulomb-fitting basis set~\cite{Wei06}. The dynamical self-energy is calculated using the contour-deformation approach~\cite{Bla11}, namely without any plasmon-pole approximation.
Structures are relaxed at the 6-31Gd PBE0 level.  BSE calculations are performed beyond the Tamm-Dancoff approximation (TDA).

\section{Results}  
We start by studying high symmetry flakes represented in Fig.~\ref{fig1}(a-d) with related data in Table~\ref{table1}.  
Besides the standard confinement-related increase of the photoemission and optical gaps with decreasing diameter $D$, we find that the HOMO and the LUMO are always doubly degenerate, consistently with the results of Refs.~\cite{Sch11,Li15}. We will return to this observation below.

\begin{table}[t]
\begin{center}
  \begin{tabular}{ccc|c|c|c}
    \hline
  $N_C$  &  $N_{tot}$ & $D$  &   gap   &  $S_{1/2}^\mathrm{dark}$      & $S^\mathrm{bright}$ ($f$)      \\
   \hline
   54  &   72  &  1.34             &   4.98   &  2.20/2.50     &   3.02 (1.06)         \\
  84  &  108  &  1.67            &   4.39   &  2.01/2.21      &  2.67 (1.44)        \\
   96  &  120  &  1.79             &   3.78   &  1.60/1.84       &  2.22 (1.24)    \\  
  114  &  144  &  1.95             &   4.06   &   1.92/2.08    &  2.45 (1.68)    \\
  \end{tabular}
  \caption{ Quasiparticle  ev$GW$  HOMO-LUMO energy gaps and BSE lowest optical absorption excitation energies (in eV) for high symmetry GQDs (Fig.~\ref{fig1}). Bright states oscillator strengths ($f$) are indicated. The average diameter $D$ (given in nm) is defined by $N_C A_C = \pi (D/2)^2$, where $N_C$ is the number of carbon atoms, $A_C = 3\sqrt{3}\, d_{CC}^2 / 4$ is the area per carbon atom with $d_{CC} = 0.142\:\mbox{nm}$. $N_{tot}$ is the total number of atoms (including hydrogen atoms). }
  \label{table1}
 \end{center}
\end{table}

Turning now to the optical properties, a remarkable feature is that the two lowest singlet excitations $S_{1/2}$  are always dark, the first bright excitations being located at higher energy (Table~\ref{table1}). 
While such properties were assigned in Ref.~\cite{Sch11} to the high symmetry ($D_{3h}$ or $D_{6h}$) of the two considered GQDs, we now show that the presence of such low-lying dark states is a much more general feature. 

We thus explore less symmetric GQDs. Starting  from the  108-atoms GQD 1(b), we  consider structures obtained by adding one  ring in all possible inequivalent positions as depicted in Fig.~\ref{fig2}(a--c), together with structures with two added rings [Fig.~\ref{fig2}(d--g)], and finally a structure with three added rings [Fig.~\ref{fig2}(h)] breaking all symmetries. In addition, an ideal and truncated rhombus-shaped GQD with armchair edges are considered [Fig.~\ref{fig2}(i,j)]. To conclude this series, we finally introduced in Fig.~\ref{fig2}(k) a structure with a very irregular edge. To facilitate the calculations on such a large number of structures, we first perform TD-DFT calculations at the PBE0/6-311Gd and TDA levels. The lowest singlet excitation ($S_1$) energy and oscillator strength  are provided below each structure. The most salient feature is that the dark nature of the lowest excitation is preserved in most cases even when all symmetries are broken [e.~g., structures~\ref{fig2}(c,g,h)]. Only the very irregular 2(k) structure presents a bright $S_1$, the dark excitation being the $S_2$ located 0.15~eV above.
%

\begin{figure*}
    \includegraphics[width= \textwidth]{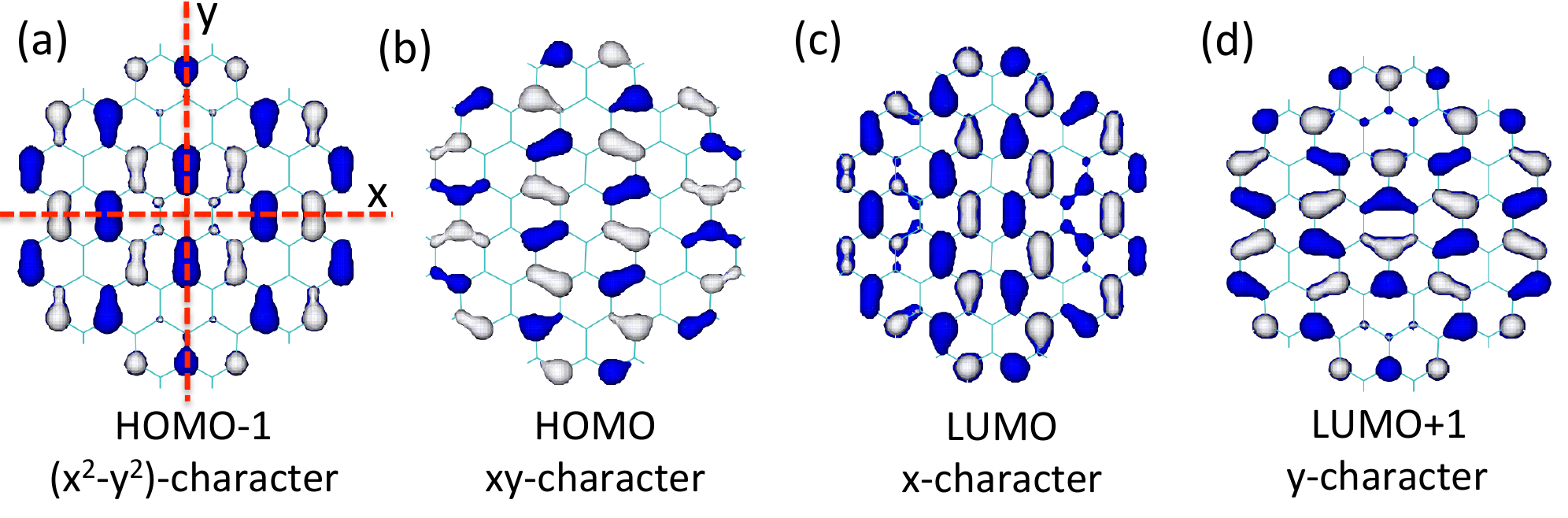}
    \caption{Isocontour representation of the PBE0 Kohn-Sham eigenstates for the frontier orbitals of the highly symmetric 1(b) structure. The dashed red lines indicate in-plane axes as a guide to the eyes.}
    \label{figwfns}
\end{figure*}

For structures shown in Figs.~\ref{fig1}(b) and~\ref{fig2}(d,h--k), we further perform full \textit{ab initio} BSE calculations which confirm the results obtained at the TD-DFT level. The BSE data are provided in Table~\ref{tab:energies} for the lowest $S_1$ singlet excitations with a more complete account in \ref{BSEeigenstates} (Table~\ref{tableS1}) for higher lying states, together with a representation of the ordering of dark and bright excitations for such systems (Fig.~\ref{fig_A1_BSE}). The most salient feature is that again in all cases the lowest singlet excitation is dark, except for the very distorted 2(k) structure for which the lowest bright excitation goes below the dark one with a 0.19~eV energy difference.

We now provide a very general rationale on why most pristine GQDs present a dark lowest energy singlet, except GQDs with  very irregular edge such as the 2(k) structure for which the $S_1$ excitation is suddenly brightened. We start by analyzing the single-electron energy levels and then proceed with the effect of the electron-hole interaction  on the  excited states, revealing the energy competition that stabilizes/destabilizes bright and dark states.


%

\begin{table*}
\begin{center}\begin{tabular}{c|ccc|c|c|ccc}
\hline
 & $\Delta_{H-1,H}$ & $\Delta_{H,L}$ & $\Delta_{L,L+1}$ & $d_{2x},\,d_{2y}$ & $d_{3x},\,d_{3y}$  &  $S_1$~energy & $f$ & $\lbrace X_i \rbrace$-weights\\
\hline
1(b) & 0.000 & 4.181 & 0.000 & $4.48,\,0.01$ & $4.48,\,0.01$       &  2.01 & 0.000 &   \\
2(d) & 0.137 & 4.020 & 0.143 & $2.31,\,-4.11$ & $2.30,\,-4.09$     &  1.86 & 0.000 &  0.48 $X_2$ + 0.48 $X_3$ \\ 
2(h) & 0.222 & 3.856 & 0.228 & $4.35,\,-2.00$ & $4.39,\,-2.03$     &  1.81 & 0.000 & 0.48 $X_2$ + 0.48 $X_3$ \\
2(j) & 0.276 & 4.295 & 0.288 & $3.54,\,2.05$ & $3.45,\,2.00$       &  2.21 & 0.000 & 0.47 $X_2$ + 0.45 $X_3$ \\
2(i) & 0.351 & 4.151 & 0.363 & $3.62,\,2.10$ & $3.67,\,2.13$       &  2.18 & 0.000 & 0.47 $X_2$ + 0.46 $X_3$ \\
2(k) & 0.632 & 3.590 & 0.633 & $-2.77,\,3.27$ & $-2.70,\,3.24$     &  1.79 & 0.711 & 0.97 $X_1$ + 0.02 $X_4$ \\
\end{tabular}\end{center}
    \caption{ \textit{Ab initio} $GW$ single-particle level spacings  (in eV) for selected structures from Figs.~\ref{fig1} and ~\ref{fig2}, ordered according to the splitting $\Delta_{H-1,H}$,
    together with the dipole matrix elements $\mathbf{d}_2$, $\mathbf{d}_3$ (in atomic units). 
    The two last columns give information on the lowest Bethe-Salpeter $S_1$ singlet eigenstates, namely energy (in eV), oscillator strength~$f$, and dominant coefficients (squared) on the $\lbrace X_i \rbrace$ transitions (see Fig.~\ref{fig3}). In the case of the 1(b) structure, the HOMO and LUMO 2-fold degeneracy hinders such a decomposition.}  
\label{tab:energies}
\end{table*}

\section{Discussion: single-particle states}

To understand the structure of single-electron orbitals, let us first consider a GQD with $C_{6v}$ symmetry (we do not exploit the full $D_{6h}$ symmetry of the structures in Fig.~\ref{fig1}). $C_{6v}$ has 4 one-dimensional irreducible representations and 2 two-dimensional ones, $E_1$ and $E_2$ represented by the functions $(x,y)\sim{E}_1$ and $(x^2-y^2,2xy)\sim{E}_2$. The four zero-energy states at the Dirac point of an infinite crystal form a representation of $C_{6v}$ which is reducible as $E_1+E_2$. 
In order to have the largest overlap with the graphene zero-energy states, the HOMO and LUMO states in a sufficiently large GQD should also correspond to $E_1$ and $E_2$ (or vice versa) and thus be doubly degenerate. The plot of the eigenstates, given in Fig.~\ref{figwfns}  for the case of the 1(b) structure, clearly confirms this analysis. Moreover, dipole transitions between $E_1$ and $E_2$ are allowed.

When the symmetry is lowered, the degeneracy is lifted. Still, in  all structures we studied, the splitting between HOMO and HOMO$-$1 or LUMO and LUMO+1 ($\Delta_{H-1,H}$ and $\Delta_{L,L+1}$, respectively)  remains small enough so that the lowest bright and dark excitations are mostly built from transitions between the HOMO$-$1, HOMO and LUMO, LUMO+1 levels. This was noted in Refs.~\cite{Sch11,Gee16} for symmetric structures, and is also confirmed by the analysis of our  BSE eigenstates (see Table~\ref{tableS1} in \ref{BSEeigenstates} and discussion below).  

Besides the overall point group symmetry, another crucial symmetry arises from the nearest-neighbour tight-binding (NNTB) model description that, in spite of its simplicity, captures many properties of graphene~\cite{CastroNeto2009}. The NNTB Hamiltonian has a chiral symmetry: it has matrix elements only between atoms from different sublattices ($A$ and $B$) of the honeycomb lattice, but not between atoms on the same sublattice. Then, for each single-electron eigenstate with an energy~$\varepsilon$ (counted from the Dirac point) and a wave function $\phi(\mathbf{r})=\phi_A(\mathbf{r})+\phi_B(\mathbf{r})$, where $\phi_A$ and $\phi_B$ have supports near $A$ and $B$ atoms, respectively, the wave function $\phi_A(\mathbf{r})-\phi_B(\mathbf{r})$ also corresponds to an eigenstate whose energy is $-\varepsilon$. The true microscopic Hamiltonian does not have the exact chiral symmetry due to second-nearest-neighbour coupling   and the variation of onsite energies for edge carbon atoms. Still, as demonstrated here below by our many-body \textit{ab initio} calculations, the chiral symmetry signature is still present. As a first illustration,  we see from Table~\ref{tab:energies} that $\Delta_{H-1,H}$ and $\Delta_{L,L+1}$ are very close for all structures. 

\begin{figure}
 \includegraphics[width=0.48\textwidth]{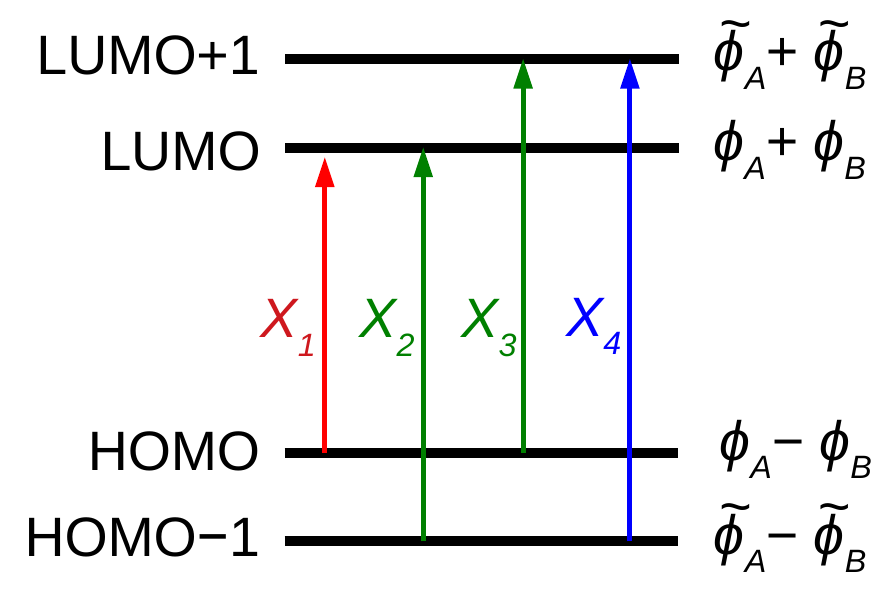}
 \caption{Electronic excitations built from the two highest occupied (HOMO${}-1$ and HOMO) electronic energy levels and symmetric lowest unoccupied (LUMO and LUMO${}+1$) energy levels with associated wavefunctions.} 
 \label{fig3}
\end{figure}

The resulting picture of the four lowest-energy electronic excitations that we denote by $\lbrace X_1,\cdots,X_4 \rbrace$ is summarized in Fig.~\ref{fig3}. Due to the chiral symmetry, we can write the single-particle wavefunctions of HOMO, LUMO and HOMO${}-1$, LUMO${}+1$ as
\begin{equation}\label{eq:wavefunctions}
\phi_{H,L} =\phi_A\mp\phi_B,\quad \phi_{H-1,L+1}=\tilde\phi_A\mp\tilde\phi_B.
\end{equation}
The chiral symmetry requires that 
excitations $X_2$ and $X_3$ are degenerate even if the GQD shape is not symmetric. Moreover, the dipole matrix elements associated with transitions $X_2,X_3$ given by:  
 \begin{eqnarray}
&&  {\bf d}_2 = \langle  \tilde{\phi}_A - \tilde{\phi}_B | {\bf r} |  {\phi}_A + {\phi}_B \rangle,\label{eq:dipole2=}\\ 
&&  {\bf d}_3 = \langle  {\phi}_A - {\phi}_B | {\bf r} |  \tilde{\phi}_A + \tilde{\phi}_B \rangle,\label{eq:dipole3=}
\end{eqnarray}
are both equal to $\langle   {\phi}_A   | {\bf r} |  \tilde{\phi}_A  \rangle - \langle  {\phi}_B   | {\bf r} |  \tilde{\phi}_B  \rangle $ 
since wavefunctions on sublattices $A$ and $B$ do not overlap. Again, in our microscopic BSE calculations not assuming the chiral symmetry, the resulting dipoles are very close (within 3$\%$; see Table~\ref{tab:energies}). As a result, the linear combination $(X_2-X_3)/\sqrt{2}$ is \emph{dark}. 
We now rationalize why this combination often becomes the lowest-energy excitation.

\begin{table*} 
\begin{center}\begin{tabular}{c|ccccc|ccccc}
\hline
 & $V_{11}^x$ & $V_{22}^x$ & $V_{33}^x$ & $V_{12}^x$ & $V_{23}^x$& $W_{11}^d$ & $W_{22}^d$ & $W_{33}^d$ & $W_{12}^d$ & $W_{23}^d$\\
\hline
1(b) &  0.400 & 0.333 & 0.331 & $10^{-4}$ & 0.313 &  2.349 & 2.313 & 2.313 & $10^{-4}$ & 0.018 \\
2(d) &  0.500 & 0.346 & 0.345 & $10^{-4}$ & 0.329 &  2.341 & 2.257 & 2.261 & $10^{-4}$ & 0.036 \\ 
2(h) &  0.494 & 0.341 & 0.341 & 0.008 & 0.325 & 2.280 & 2.231 & 2.235 & $-0.007$ & 0.032 \\
2(j) &  0.435 & 0.232 & 0.234 & $10^{-5}$ & 0.216 &  2.331 & 2.273 & 2.276 & $10^{-5}$ & 0.018\\
2(i) &  0.448 & 0.220 & 0.225 & $10^{-4}$ & 0.206 &  2.271 & 2.226 & 2.242 & $10^{-5}$ & 0.011 \\
2(k) &  0.471 & 0.317 & 0.317 & $-0.033$ & 0.304 &  2.283 & 2.201 & 2.207 & $-0.018$ & 0.077 \\
\end{tabular}\end{center}
    \caption{ \textit{Ab initio}  Bethe-Salpeter matrix elements (in eV) for selected structures from Figs.~\ref{fig1} and ~\ref{fig2}, ordered according to the splitting $\Delta_{H-1,H}$. }  
\label{tab:Coulomb}
\end{table*}

\section{Discussion: electron-hole interactions}
\label{sec:eh}

The dominant resonant part of the 2-body electron-hole BSE Hamiltonian in the  $\lbrace X_1,\ldots,X_4 \rbrace$ basis reads: 
\begin{equation}
H^{BSE}_{ia,jb} = (\varepsilon_a-\varepsilon_i)\, \delta_{ij} \delta_{ab} + 2 V_{ia,bj}^x  - W_{ij,ab}^d,
\end{equation}
where $(i,j)$/$(a,b)$ label occupied/empty single-particle levels, respectively, and:
\begin{eqnarray*}
&&V^x_{ia,bj}=\int\phi_i(\mathbf{r}) \phi_a(\mathbf{r}) V(\mathbf{r}-\mathbf{r}') \phi_b(\mathbf{r}') \phi_j(\mathbf{r}')\,d\mathbf{r}\, d\mathbf{r}', \\
&&W_{ij,ab}^d=\int\phi_i(\mathbf{r})\phi_j(\mathbf{r}) W(\mathbf{r},\mathbf{r}') \phi_a(\mathbf{r}')\phi_b(\mathbf{r}') \, \rmd\mathbf{r}\, \rmd\mathbf{r}',
\end{eqnarray*}
are matrix elements of the bare and screened Coulomb potential, respectively.
Selected \textit{ab initio} matrix elements for the considered structures are given in Table~\ref{tab:Coulomb}. Their inspection reveals several common properties.

(i)~The diagonal elements $W^d_{\alpha\alpha}$ are by far dominant and have close values. 
This is a consequence of the long-range nature of the Coulomb interaction and orthogonality of the single-electron wave functions: if one takes the limit of the infinite-range interaction for $V$ and $W$, namely   $V(\mathbf{r}-\mathbf{r}'),\,W(\mathbf{r},\mathbf{r}')\to\mathrm{const}$, the integrals over $\mathbf{r}$ and $\mathbf{r}'$ separate and vanish by orthogonality everywhere except $W^d_{\alpha\alpha}$ which are all equal.

(ii)~Both $V^x_{\alpha\beta}$ and $W^d_{\alpha\beta}$ are small if $\alpha\in\{1,4\}$ and $\beta\in\{2,3\}$. In fact,  they must vanish for structures 1(b), 2(d,i,j) because of the remaining mirror symmetry~\cite{precision}.  Indeed, HOMO and HOMO${}-1$ must have different parity with respect to this mirror reflection, and the same holds for LUMO and LUMO${}+1$; this follows from the reduction of the two-dimensional $E_1$ and $E_2$ representations of the $C_{6v}$ group to a simple mirror one. Since the Coulomb potential is even, the corresponding integrals must vanish. For GQDs which do not have exact mirror symmetry, these   integrals  remain small.

(iii) The diagonal $V_{\alpha\alpha}^x>0$, because the Coulomb potential $V(\mathbf{r}-\mathbf{r}')$ is positive in the operator sense.

(iv) $V^x_{22}\approx V^x_{33}\approx V^{x}_{23}$. In fact, if one uses the representation~(\ref{eq:wavefunctions}) of the single-particle wave functions and neglects the overlap between $A$ and $B$ components, all three integrals become $\langle\phi_A\tilde\phi_A-\phi_B\tilde\phi_B|V|\phi_A\tilde\phi_A-\phi_B\tilde\phi_B\rangle$, another signature of the underlying chiral symmetry.

(v) $W^d_{23}\ll V^{x}_{23}$. In the representation of Eq.~(\ref{eq:wavefunctions}), these matrix elements have the form   $\langle\phi_A\tilde\phi_A\pm\phi_B\tilde\phi_B|V|\phi_A\tilde\phi_A\pm\phi_B\tilde\phi_B\rangle$ with the $+/-$ sign corresponding to $W^d_{23}$/$V^{x}_{23}$, respectively. 
 $V^x_{23}$~is mostly the dipole-dipole interaction between $\mathbf{d}_2$ and $\mathbf{d}_3$. In contrast, for the symmetric 1(b) and 2(i) structures, the dipole moment of the $\phi_H \phi_{H-1}$ and $\phi_L \phi_{L+1}$ co-densities vanishes by symmetry, so that $W^d_{23}$ stems from a weaker quadrupole-quadrupole interaction. For  other structures with reduced symmetry, the codensity dipoles remain significantly smaller than~$\mathbf{d}_{2/3}$.

Moreover, for the $\phi_H \phi_{H-1}  =  \phi_L \phi_{L+1}  =   \phi_A\tilde\phi_A + \phi_B\tilde\phi_B$  co-densities we find a strong sign alternation on adjacent A/B sites, which persists even in the absence of  spatial (point group) symmetries. This leads to a further suppression of $W^d_{23}$ due to the  smooth long-range nature of the Coulomb potential, so that for rapidly oscillating functions the integral is small. We rationalize this observation in \ref{nntbmodel}.

These properties determine the energy ordering of the optical excitations. The large diagonal matrix elements $W^d_{\alpha\alpha}$ provide an overall shift without affecting much the ordering, by virtue of property~(i). Due to property~(ii), the $\lbrace X_2,X_3 \rbrace$ and $\lbrace X_1,X_4 \rbrace$ blocks do not couple and can be analyzed separately. This is confirmed by the stable and dominant weights of the BSE lowest dark states on the $X_{2/3}$ transitions (see Table~\ref{tab:energies}) irrespective of the $\Delta_{H-1,H}$ splitting. Similarly, the lowest BSE bright states are   built dominantly from $X_1$ and $X_4$ contributions (see structure 2(k) in Table~\ref{tab:energies} and Table~\ref{tableS1} in \ref{BSEeigenstates}). In the $\lbrace  X_2,X_3 \rbrace$ block, the degeneracy is lifted essentially by $V_{23}^x$ [property~(v)] which pushes up in energy  the bright combination $(X_2+X_3)/\sqrt{2}$ by $4V^x_{23}$ and leaves the energy of the dark combination unaffected, due to properties~(iii,iv). In the $\lbrace X_1,X_4 \rbrace$ sector, the main effect is to push up $X_1$ by $2V_{11}^x$ [property~(iii)]; the mixture with $X_4$ and the corresponding level repulsion is less important since $X_1$ and $X_4$ were split in the very beginning because of the single-particle energy difference between $\Delta_{HL}$ and $\Delta_{H-1,L+1}$ when the degeneracies of HOMO and LUMO are broken. 
In particular, for structure 2(k) with largest $\Delta_{H-1,H}$ splitting, the lowest bright excitation is dominated by the $X_1$ HOMO-LUMO transition (see Table~\ref{tab:energies}). In practice, $(X_2+X_3)/\sqrt{2}$ and $X_4$ are  high enough in energy, so that they mix with higher excitations.   

Overall, the difference in energy between the dark $(X_2-X_3)/\sqrt{2}$  and bright  $X_1$ excitations read:
$$
E^{\mathrm{dark}}-E^{\mathrm{bright}} = \Delta_{H-1,H} + W_{23}^d +  W_{11}^d - W_{22}^d  - 2V_{11}^x 
$$ 
with $W_{23}^d \simeq 0$ and $(W_{11}^d - W_{22}^d) \simeq 0$.
As a result, the relative position of the lowest $X_1$-like bright excitation and the ($X_2-X_3$) dark combination 
depends on the competition between the energy splitting $\Delta_{H-1,H}\approx\Delta_{L-1,L}$ 
pushing the dark state up, and the diagonal exchange integral $2V_{11}^x$ that pushes $X_1$ up. Thus, for high-symmetry GQDs with degenerate HOMOs and LUMOs, the lowest excitation is always dark. 
Upon breaking the point-group symmetry, the level splitting $\Delta_{H-1,H}\approx\Delta_{L-1,L}$ increases as the shape of the dot becomes more irregular. Still, the lowest excitation remains the dark one as long as the splitting remains small compared to the energy scale which can be roughly estimated as $2V_{11}^x$.

The non-zero contributions from the $W_{23}^d$ and $(W_{11}^d - W_{22}^d)$ terms, due to deviations of the BSE \textit{ab initio} Hamiltonian from the NNTB model, the small coupling of the $X_1$ state with the higher-lying $X_4$ contribution, and the non-resonant contributions in our full (beyond TDA) BSE calculations, explain that the criterion on the sign of ($\Delta_{H-1,H}  - 2V_{11}^x$) is not strictly quantitative. This is the reason why the structure 2(k) has a bright ground state while the splitting is still smaller that $2V_{11}^x$.


These findings are reminiscent of carbon nanotubes with the presence of low-lying dark excitations related to the mixing and stabilization by Coulomb interaction of degenerate single-particle transitions~\cite{Mazumdar_2004}. Due to reduced confinement and enhanced screening as compared to GQDs, differences of energy between bright and dark excitonic states in nanotubes were found to be of the order of a few  meV~\cite{Mortimer_2007}, namely, much smaller than what we observe here.  Similar effects have been revealed in the case of nanoribbons  but with a strong dependence on the edge structure (the ``geometric chirality'', which must not be confused with the chiral symmetry we consider in the present paper) that governs the presence of degenerate single-particle transitions at low energy~\cite{Molinari_2008}. We emphasize however that what we addressed here is the case of finite size systems with irregular edges, revealing the properties leading to the ``protection'' of the low-lying dark states upon breaking of the ideal geometry and lifting of the single-particle transition degeneracies.

\section{Effect of functionalization}
To confirm the importance of the chiral symmetry, we finally consider the experimental observation that most photoluminescent GQDs are functionalized with nitrogen or oxygen-rich side groups \cite{Pan10,Li10,Li11,Zhu11,Shen12,Peng12,Lim15,Ding16}. Taking as a paradigmatic example the 108-atoms GQD (Fig.~\ref{fig1}b), we first replace four passivating H atoms by carboxyl -COOH groups, leaving all carbon atoms in an sp$^{2}$ configuration. Such a functionalization hardly lifts the HOMO and LUMO degeneracy and leaves the 2 lowest excitation dark. On the contrary, adding the carboxyl groups to already H-functionalized C atoms, namely, creating $sp^3$ edge carbon atoms [see Fig.~\ref{fig1}(e)], significantly increases the $S_1$ oscillator strength from zero to 0.25.  
Our conclusion is that the asymmetric functionalization (one H atom, one carboxyl group) significantly distorts the flake that starts deviating from its ideal $sp^2$ planar geometry, breaking the underlying conditions for chiral symmetry.   Replacing the COOH group by an H atom that we fully relax  but keeping  all C atoms frozen to their distorted geometry, we find that the $S_1$ excitation remains bright with an energy and oscillator strength that hardly changes. 

Exploring all possible ways to  distort planar GQDs by functionalization (see e.g. Ref.~\cite{Cocchi2012}) is beyond the scope of the present study. We can conclude however that \textit{sp}$^3$ functionalization provides an important pathway to increase emission from GQDs by brightening the lowest singlet excitations populated by Kasha's rule after relaxation of hot electrons. We emphasize however that such a behaviour does not directly involve the electronic properties of the side groups that just serve as a mean to induce a structural distortion breaking the chiral symmetry. 
The $sp^3$ character of edge atoms was revealed explicitly in a few studies by XPS measurements~\cite{Tang12,Ding16}.  Such a picture allows to make a possible connection between  GQDs and graphene oxide that can be described as \textit{sp}$^2$ islands in an $sp^3$ matrix \cite{Eda10,Loh10}.

\section{Structural relaxation in the excited state}

To conclude this exploration, we now study the effect of relaxation in the excited state. We turn again to TD-DFT calculations at the PBE0/6-311Gd level within the Tamm-Dancoff approximation for sake
of numerical efficiency. The lack of analytic forces in the BSE formalism precludes relaxations in the excited state.  The good agreement between TD-PBE0 and BSE data concerning the energy and nature (bright or dark) of the lowest absorption singlet states,  allows to conclude that both TD-DFT and BSE convey the same physical picture. This is an indirect signature that the systems we study are not presenting charge-transfer states that would result in difficulties for TD-PBE0.  

Upon relaxation in the $S_1$ state, the GQD-72 (Fig.~1a) lowest excitation energy is redshifted from 2.33 eV to 2.26 eV (TD-DFT values), but the lowest $S_1$ state remains completely dark. Similarly, the GQD-108 (Fig.~1b) singlet absorption onset $S_1$ is redshifted upon relaxation from 2.24 eV to 2.07 eV remaining completely dark. Clearly, relaxation in the excited state induces a marginal effect on the emission energy for such pristine GQDs. Turning now to the structure  with \textit{sp}$^2$-preserving functionalization, we find again that relaxation shifts the $S_1$ energy by about 0.06 eV, from 2.07 eV to 2.01 eV, but without changing its dark state nature. 

 Finally, relaxation of the \textit{sp}$^3$-functionalized (Fig.~1e) system results in a 0.1  eV redshift of the lowest excited state (from 2.12 to 2.02 eV), its oscillator
strength changing from 0.31 to 0.43 (TD-PBE0/TDA values). These selected results clearly indicate that  relaxation in the $S_1$ excited state does not  affect the conclusions drawn here above in the study of the lowest absorption features.

\section{Conclusions}
In conclusion, we demonstrated that pristine GQDs present in general a dark lowest singlet excitation as a result of the underlying chiral symmetry, irrespective of the overall geometrical point-group symmetry. 
The relation between electron and hole wavefunctions on sublattices A and B, as induced by the local chiral symmetry in graphene, ensures that there is always a dark combination of the (HOMO)$\rightarrow$(LUMO+1) and (HOMO-1)$\rightarrow$LUMO transitions that lies in general lower in energy than the bright HOMO$\rightarrow$LUMO excitation destabilized by exchange Coulomb interaction. Only very distorted structures, characterized by a large HOMO-(HOMO-1)  or equivalent LUMO-(LUMO+1)  splitting destabilizing the dark combination, can bring the bright HOMO-LUMO excitation lower in energy.  

While extensive stability calculations may provide insights on the likeliness of very irregular edge structures given the synthesis conditions~\cite{Was08}, we conclude that a very large fraction of pristine GQDs, including GQDs without any overall point-group symmetry,  present a lowest dark singlet quenching photoluminescence. Such a conclusion is not affected by the weak structural relaxation in the excited state.  Our results indicate that   \textit{sp}$^3$ edge functionalization is extremely efficient in  switching the photoluminescent properties of GQDs. We observe however that it is the breaking of the underlying chiral symmetry, and not necessarily the occurrence of charge-transfer core-to-edge optical excitations, that plays a significant role.  

Our results do not exclude other pathways to enhancing photoluminescence. Consistently with early reports on the effect of doping in nanotubes~\cite{Haru_2009}, the doping of GQDs \cite{Sk14,Niu16,Noo16,Fen18,Kadian_2019} also stands as an efficient way to enhance photoluminescence. Following the well known case of graphene~\cite{Zhou2007}, the interaction with a substrate may also affect electronic properties close to the gap, with an effect on the photoluminescence related to the interaction strength. Clearly,  doping or charge-transfer from a substrate,  introduction of  in-gap defect-induced levels or  associated structural distortions, can break the pristine GQDs   electron-hole symmetry.   Finally, functional groups optically active in the GQD optical gap can also stand as a way to bypass the presence of these low-lying dark states.

The experimental characterization of such dark states  would require the synthesis of undoped GQDs with controled planar shape and the verification that such systems are poorly photoluminescent. We do not exclude that this has been observed but for obvious reasons  emphasis was put on systems with good photoluminescent properties. 
As a suggestion, comparing functional groups known to favour $sp^2$ versus $sp^3$ edge functionalization should allow to relate the photoluminescent properties to the GQD planarity, as $sp^3$ functionalization is likely to  distort significantly the GQD.

\ack
The authors acknowledge support from the French GENCI supercomputing facilities under contract A0030910016. 

\section*{References}

\bibliographystyle{iopart-num}
\bibliography{biblio}

\begin{thebibliography}{10}
\bibitem{precision}
The values $10^{-4}-10^{-5}$ correspond to the fact that we did not impose a
  strict constraint on symmetries during the structural relaxation.
\end{thebibliography}

\newpage
\appendix

\section{Analysis of the Bethe-Salpeter excitations}
\label{BSEeigenstates}

\begin{figure}[b]
 \includegraphics[width=0.48\textwidth]{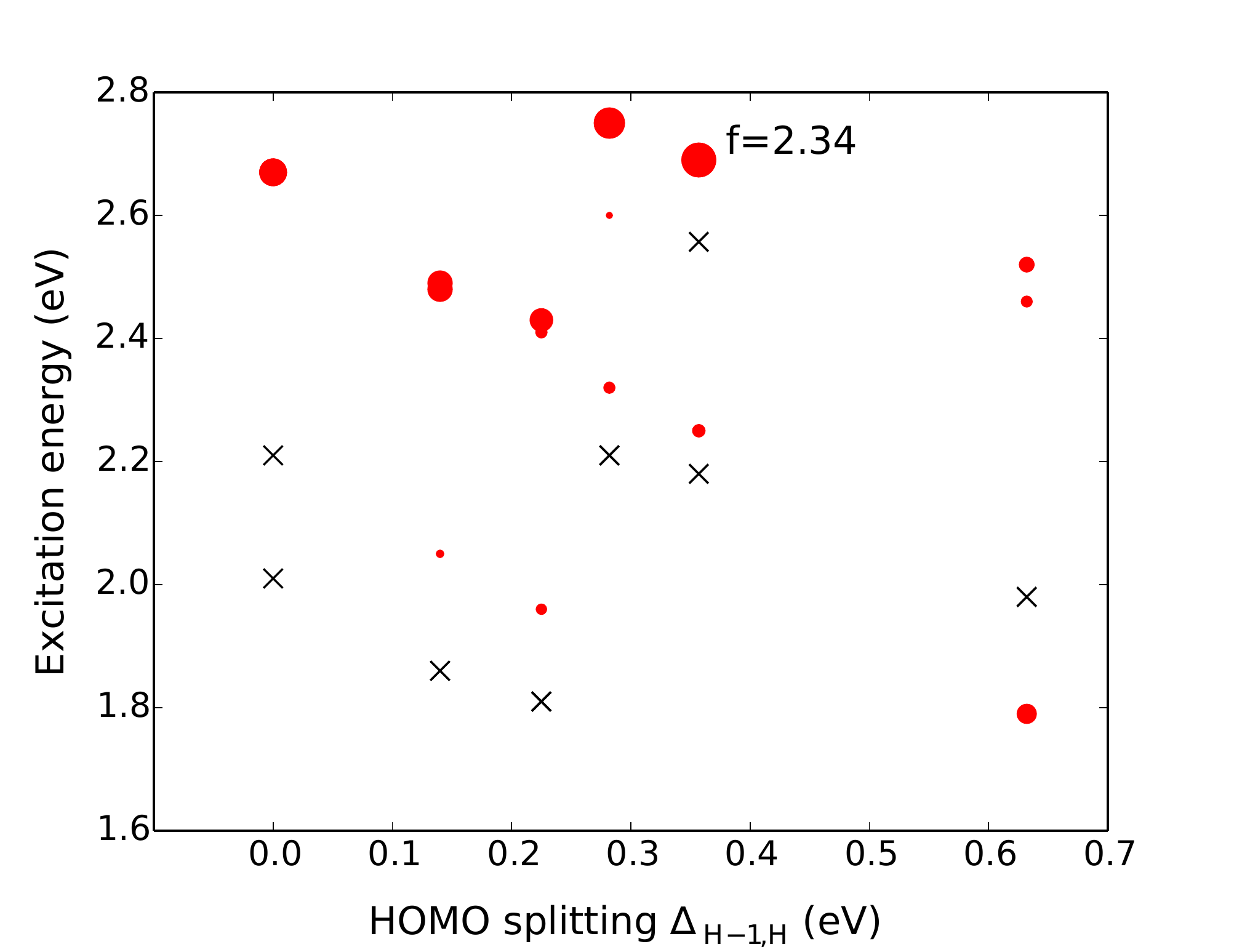}
 \caption{ Representation of the data of Table A1 : lowest singlet excitation energies for structures 1(b) and 2(d,h,i-k). The systems are ordered according to the $\Delta_{H-1,H}$ HOMO splitting. Black crosses represent dark states; bright excitations are represented by red circles with a radius proportional to the oscillator strength. The largest (f) oscillator strength is given to set the circle size scale. } 
 \label{fig_A1_BSE}
\end{figure}

A more detailed account of the BSE excitations characteristics is provided in  Table~\ref{tableS1}.
Besides the excitation energy, the associated oscillator strength (in parenthesis) and leading weights of the 2-body eigenstates on the low-energy $\lbrace X_1,\ldots,X_4 \rbrace$ transitions, but also on higher $X_i$ transitions for  excitations above the lowest dark and bright combinations, are provided. The Bethe-Salpeter eigenstates read:
\begin{eqnarray}
&&\psi^{BSE}_{\lambda}( {\bf r}_e , {\bf r}_h ) = \sum_{ia} A_{\lambda}(ia) \phi_i({\bf r}_h) \phi_a({\bf r}_e)   +{}\nonumber\\
&&\hspace*{2.5cm}{}+ 
\sum_{ia} B_{\lambda}(ia) \phi_i({\bf r}_e) \phi_a({\bf r}_h),
\end{eqnarray}
where $(i,a)$ label occupied/empty energy levels and $\lambda$ the Bethe-Salpeter excitations with increasing energy.  Table~\ref{tableS1} provides the leading $|A_{\lambda}(ia)|^2$ coefficients. The $B_{\lambda}(ia)$ coefficients represent de-excitations when going beyond the Tamm-Dancoff approximation and are much smaller in magnitude. We use the notations of Fig.~\ref{fig3}: $X_1$ represents the HOMO to LUMO transition, $X_2$ the HOMO$-1$ to LUMO transition, $X_3$ HOMO to LUMO$+1$, and $X_4$ HOMO$-1$ to LUMO$+1$. In addition, we introduce $X_5$ (HOMO$-2$ to LUMO), $X_6$ (HOMO to LUMO$+2$) and $X_7$ (HOMO to LUMO$+3$). All dark and bright states compiled in Table~\ref{tableS1}  are represented in Fig.~\ref{fig_A1_BSE}.

\begin{table*}[t]
 \begin{center} \begin{tabular}{c|c|c|c|c }
   \hline
         &           $S_1$                          &             $S_2$                     &       $S_3$                             &            $S_4$                                         \\    \hline       
1(b)     &    \bf{      2.01 (0.000)      }                  &         2.21 (0.000)                   &        2.67  (1.439)                       &          2.67  (1.438)                                \\ \hline
2(d)     &     \bf{      1.86 (0.000)     }                  &        2.05  (0.091)                   &         2.48  (1.181)                      &   2.49  (1.175)                                       \\
          &  \bf{  0.48  $X_2$ +  0.48   $X_3$   }       &         0.72 $X_1$ + 0.26 $X_4$          &        0.44  $X_2$ + 0.44 $X_3$             &   0.68 $X_4$ + 0.25 $X_1$                                  \\ \hline
2(h)       &  \bf{       1.81  (0.000)     }                  &        1.96  (0.188)                   &       2.41 (0.225)                        &  2.43 (1.015)                                          \\
           &  \bf{        0.48 $X_2$ + 0.48 $X_3$  }                &      0.81 $X_1$ + 0.18 $X_4$              &  0.46 $X_6$ + 0.16 $X_7$   &  0.34 $X_2$ + 0.35 $X_3$                                \\ \hline
 2(j)       &    \bf{      2.21  (0.000)       }               &    2.32 (0.220)                         &  2.60  (0.056)                             &  2.75   (1.877)                                      \\
           &  \bf{ 0.47 $X_2$ + 0.45 $X_3$     }                 &   0.83  $X_1$ + 0.14 $X_4$               &   0.51 $X_6$ + 0.23 $X_5$    &   0.48 $X_2$ + 0.49 $X_3$                               \\ \hline
 2(i)      &  \bf{ 2.18 (0.000) }                       &    2.25 (0.281)                         &      2.557 (0.000)                             &  2.69 (2.344)                                         \\
          & \bf{ 0.47 $X_2$ + 0.46 $X_3$ }              &   0.86 $X_1$ + 0.11 $X_4$                &     0.48 $X_6$ + 0.24 $X_5$        &  0.49 $X_2$ + 0.50 $X_3$                                 \\ \hline
2(k)      &           1.79  (0.711)                      &     \bf{ 1.98  (0.000)   }                  &     2.46  (0.211)                          &  2.52  (0.411)                                          \\
          &          0.97  $X_1$  + 0.02 $X_4$             &     \bf{  0.45 $X_2$ + 0.46 $X_3$   }      &  0.60 $X_7$ + 0.13 $X_3$             &   0.19 $X_2$ + 0.16 $X_3$                                \\ \hline
     \end{tabular}\end{center}
  \caption{ Details on the low-energy Bethe-Salpeter excitations: energy in eV (oscillator strength) and composition, for structures 1(b) and 2(d,h,i,j,k) as indexed in the main text. The dark excitations resulting from a balanced combination of the ($X_2$) and ($X_3$) transitions are highlighted in bold (see the definition of $X_i$ transitions in Fig.~\ref{fig3}). The composition cannot be determined unambiguously for the first structure because of the doubly degenerate HOMO and LUMO. } 
  \label{tableS1}
\end{table*}

\section{Nearest-neighbor tight-binding results}
\label{nntbmodel}

The nearest-neighbor tight-binding model is less precise quantitatively than the \textit{ab initio} calculations, but it is very useful when one needs a qualitative insight into the properties of the wave functions. Here we give the results for the structure from Fig.~2(i) of the main text, a rhombus with armchair edges, possessing the $D_{2h}$ symmetry. Since edge passivation by hydrogen was not included in the tight-binding calculation, no quantitative comparison to the \textit{ab initio} results can be made.

\begin{figure*}
\begin{center}
    \includegraphics[width=0.42\textwidth]{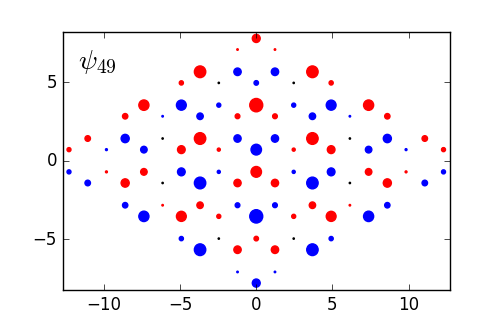}
    \includegraphics[width=0.42\textwidth]{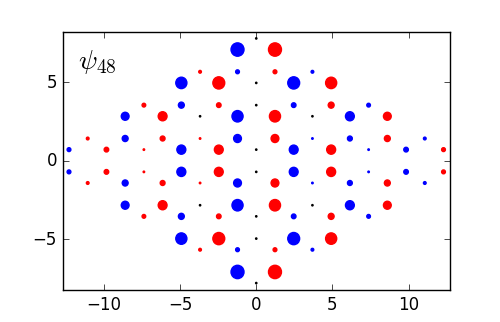}\\
    \includegraphics[width=0.42\textwidth]{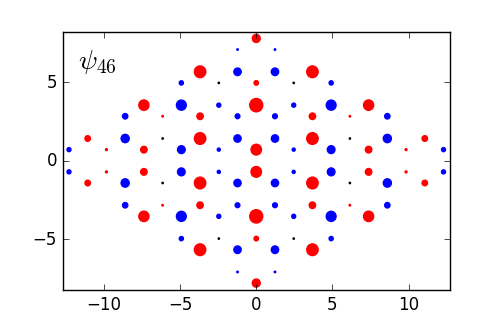}
    \includegraphics[width=0.42\textwidth]{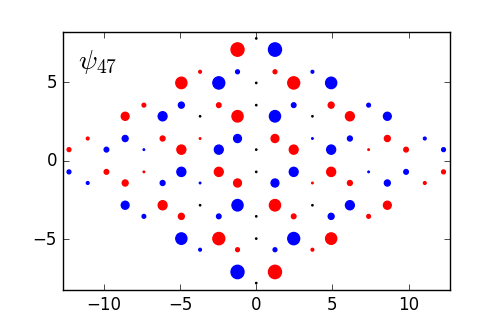}
    \end{center}
    \caption{Tight-binding single-particle wave functions $\psi_{49}$ (LUMO$+1$, energy 1.25~eV, symmetry~$\sim{y}$), $\psi_{48}$ (LUMO, energy 0.96~eV, symmetry~$\sim{x}$), $\psi_{47}$ (HOMO, energy $-0.96$~eV, symmetry~$\sim{xy}$), $\psi_{46}$ (HOMO$-1$, energy $-1.25$~eV, symmetry $\sim{x^2-y^2}$) for structure 2(i). The wave functions are real, the color of each circle indicates the sign, while the circle radius represents the absolute value. The carbon atom positions are in angstroms.}
    \label{fig:wave_functions}
\end{figure*}

Diagonalization of the tight-binding Hamiltonian, determined by the nearest-neighbor matrix element $\gamma_0=3.3\:\mbox{eV}$ yields 96 single-particle eigenstates $\psi_0,\psi_1,\ldots,\psi_{95}$, which we order according to their energies. The important orbitals HOMO$-1$, HOMO, LUMO and LUMO$+1$ ($\psi_{46}$, $\psi_{47}$, $\psi_{48}$ and $\psi_{49}$, respectively) are shown in Fig.~\ref{fig:wave_functions}.

To see the effects of the wave function structure on the Coulomb integrals, for both direct and exchange parts, we use the unscreened Coulomb interaction
\begin{equation}
    V_{nm}=\left\{\begin{array}{ll}
    \displaystyle\frac{e^2}{|\mathbf{r}_n-\mathbf{r}_m|}, & n\neq m,\\
    U, & n=m,
    \end{array}\right.
\end{equation}
where $n,m$ label the sites of the honeycomb lattice. The on-site repulsion $U=15.75\:\mbox{eV}$ was obtained by evaluating the Coulomb integral on the microscopic $p_z$ orbital of a carbon atom. With this value, the matrix $V_{nm}$ is positive-definite for any GQD shape (indeed, the minimal value of $U$ required to make the matrix $V_{nm}$ positive-definite for an infinite honeycomb lattice with nearest-neighbor distance 1.42~{\AA} is $U_\mathrm{min}=15.64\:\mbox{eV}$). The resulting exchange and direct integrals $V_{\alpha\beta}^x$, $V_{\alpha\beta}^d$ in the basis $X_1,X_2,X_3,X_4$ are given by (in~eV)
\begin{eqnarray}
V^x=\left(\begin{array}{cccc}
0.447  & 0.          & 0.          & -0.315 \\
0.     & 0.244 & 0.244 & 0.         \\
0.     & 0.244 & 0.244 & 0.         \\
-0.315 & 0.          & 0.          & 0.311
\end{array}\right),\\
V^d=\left(\begin{array}{cccc}
 2.660  & 0.     & 0.     & 0.023 \\
 0.     & 2.575  & 0.023  & 0.         \\
 0.     & 0.023  & 2.575  & 0.         \\
 0.023  & 0.     & 0.     & 2.670
\end{array}\right).
\end{eqnarray}
They have the same proprties (i)--(v) as the \textit{ab initio} Coulomb integrals, discussed in Sec.~\ref{sec:eh}.

\begin{figure*}
\begin{center}
    \includegraphics[width=0.47\textwidth]{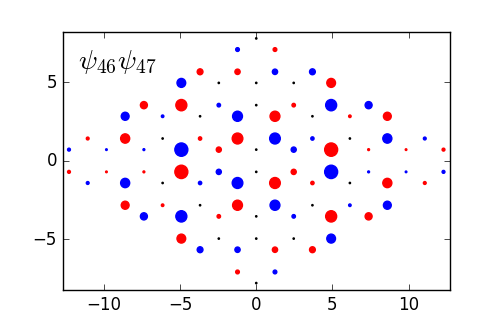}
    \includegraphics[width=0.47\textwidth]{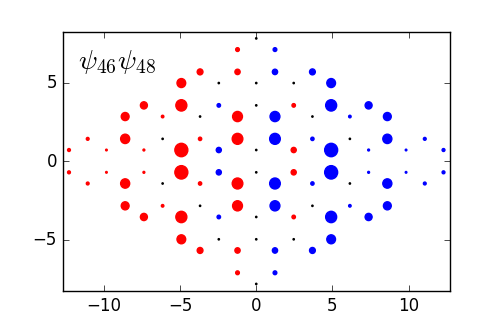}
    \end{center}
    \caption{Co-densities $\psi_{46}\psi_{47}$ (left) and $\psi_{46}\psi_{48}$ (right). The color indicates the sign while the dot size represents the absolute value. The positions are in angstroms. Note the sign alternation on neighbouring sites for $\psi_{46}\psi_{47}$. }
    \label{fig:codensities}
\end{figure*}

In Fig.~\ref{fig:codensities} we show the co-densities $\psi_{46}\psi_{47}$ and $\psi_{46}\psi_{48}$, responsible for $V^d_{23}$ and $V^x_{23}$, respectively. Indeed, since $\psi_{46}\psi_{47}=\psi_{48}\psi_{49}$ and $\psi_{46}\psi_{48}=\psi_{47}\psi_{49}$ exactly, due to the chiral symmetry of the nearest-neighbor tight-binding model, we have
\begin{eqnarray}
&&V^d_{23}=\langle\psi_{46}\psi_{47}|V|\psi_{46}\psi_{47}\rangle,\\
&&V^x_{23}=\langle\psi_{46}\psi_{48}|V|\psi_{46}\psi_{48}\rangle.
\end{eqnarray}
The symmetries $\psi_{46}\psi_{47}\sim{xy}$ (quadrupole), $\psi_{46}\psi_{48}\sim{x}$ (dipole) follow straightforwardly from the symmetries of the corresponding wave functions. Another striking difference between the two co-densities is that $\psi_{46}\psi_{47}$ exhibits pronounced sign alternation on neighboring sites, while $\psi_{46}\psi_{48}$ represents a rather smooth spatial dependence.

To be more quantitative about the smoothness, and to understand the origin of the sign alternation in $\psi_{46}\psi_{47}$, we introduce the envelope functions. Namely, separating a real wave function $\phi(\mathbf{r})$ into sublattice components, $\phi(\mathbf{r})=\phi_A(\mathbf{r})+\phi_B(\mathbf{r})$, we can represent each component $\phi_{A,B}(\mathbf{r})$ as
\begin{equation}
\phi_{A,B}(\mathbf{r})=\Psi_{A,B}(\mathbf{r})\,e^{\rmi\mathbf{K}\mathbf{r}}+\Psi_{A,B}^*(\mathbf{r})\,e^{-\rmi\mathbf{K}\mathbf{r}},
\end{equation}
where $\Psi_{A,B}(\mathbf{r})$ are the smooth envelopes and $\pm\mathbf{K}$ are the inequivalent corners of the first Brillouin zone. If the GQD shape is invariant under a reflection $\mathbf{r}\mapsto\mathcal{R}_{AB}\mathbf{r}$ which interchanges the two sublattices but leaves invariant the $\pm\mathbf{K}$ points (for the rhombic shape considered here, it is $y\to-y$), the wave function $\phi(\mathbf{r})$ is either even or odd, $\phi(\mathbf{r})=\pm\phi(\mathcal{R}_{AB}\mathbf{r})$. Then, the envelope functions obey $\Psi_{A,B}(\mathcal{R}_{AB}\mathbf{r})=\pm\Psi_{B,A}(\mathbf{r})$. If the GQD shape is invariant under a reflection $\mathbf{r}\mapsto\mathcal{R}_{\pm\mathbf{K}}\mathbf{r}$ which preserves the two sublattices but flips the $\pm\mathbf{K}$ points (for the rhombic shape considered here, it is $x\to-x$), the envelope functions obey $\Psi_{A,B}(\mathcal{R}_{\pm\mathbf{K}}\mathbf{r})=\pm\Psi_{A,B}^*(\mathbf{r})$.

\begin{figure*}
\centering
        \includegraphics[width=17.cm]{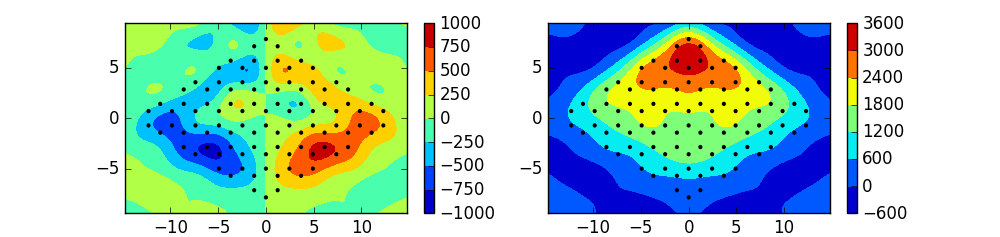}
        \vspace{-0.0cm}\\
        {\large $\Re\Psi_{47,A}$} \hspace{6cm} {\large $\Im\Psi_{47,A}$}
        \vspace{0.cm}\\
        \includegraphics[width=17.cm]{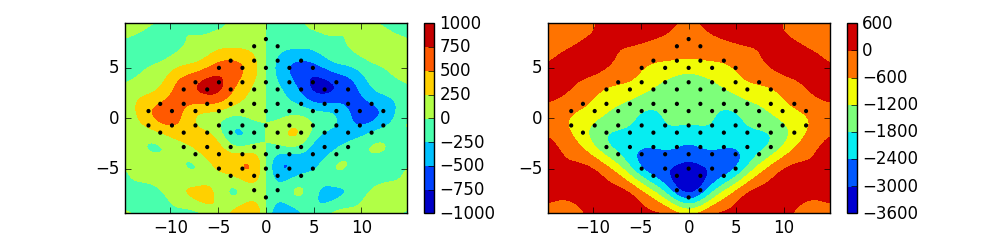}
        \vspace{-0.0cm}\\
        {\large $\Re\Psi_{47,B}$} \hspace{6cm} {\large $\Im\Psi_{47,B}$}
        \vspace{0.cm}\\
    \caption{Real and imaginary parts of the envelope functions $\Psi_A(\mathbf{r})$ and $\Psi_B(\mathbf{r})$ for HOMO, in arbitrary units.}
    \label{fig:RHOMBenvelopesReIm47}
\end{figure*}

\begin{figure*}
\centering
        \includegraphics[width=17.cm]{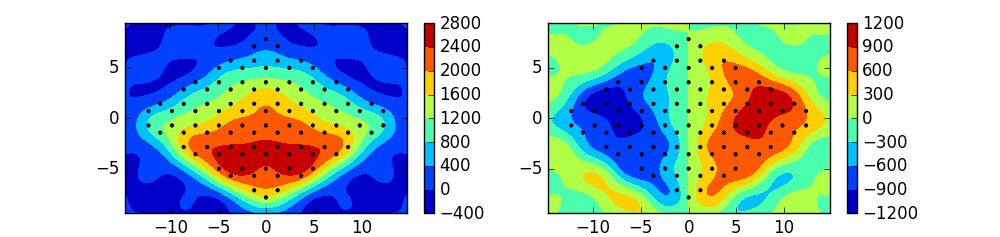}
        \vspace{-0.0cm}\\
        {\large $\Re\Psi_{46,A}$} \hspace{6cm} {\large $\Im\Psi_{46,A}$}
        \vspace{0.cm}\\
        \includegraphics[width=17.cm]{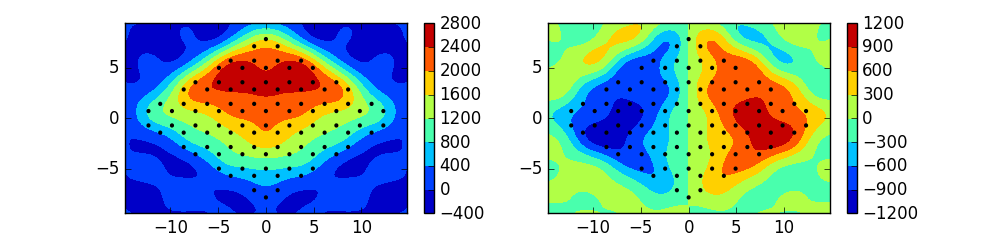}
        \vspace{-0.0cm}\\
        {\large $\Re\Psi_{46,B}$} \hspace{6cm} {\large $\Im\Psi_{46,B}$}
    \caption{Real and imaginary parts of the envelope functions $\Psi_A(\mathbf{r})$ and $\Psi_B(\mathbf{r})$ for HOMO$-1$, in arbitrary units.}
    \label{fig:RHOMBenvelopesReIm46}
\end{figure*}


The Fourier transform of each sublattice component for all four orbitals in Fig.~\ref{fig:wave_functions} is strongly peaked around $\pm\mathbf{K}$. The inverse Fourier transform of the peak around $\mathbf{K}$ determines the smooth envelope functions $\Psi_{A}(\mathbf{r})$, $\Psi_{B}(\mathbf{r})$. The complex envelope functions for states 46 and 47 are shown in Figs.~\ref{fig:RHOMBenvelopesReIm47} and \ref{fig:RHOMBenvelopesReIm46}  (real and imaginary parts). 
Due to the chiral symmetry of the nearest-neighbor tight-binding model, $\Psi_{48,A}=\Psi_{47,A}$, $\Psi_{48,B}=-\Psi_{47,B}$,
$\Psi_{49,A}=\Psi_{46,A}$, $\Psi_{49,B}=-\Psi_{46,B}$. Let us denote $\Psi_{48,A}=\Psi_{47,A}=\Psi_A$, $\Psi_{48,B}=-\Psi_{47,B}=\Psi_B$, $\Psi_{49,A}=\Psi_{46,A}=\tilde\Psi_A$, $\Psi_{49,B}=-\Psi_{46,B}=\tilde\Psi_B$.

\begin{figure}
    \includegraphics[width=0.47\textwidth]{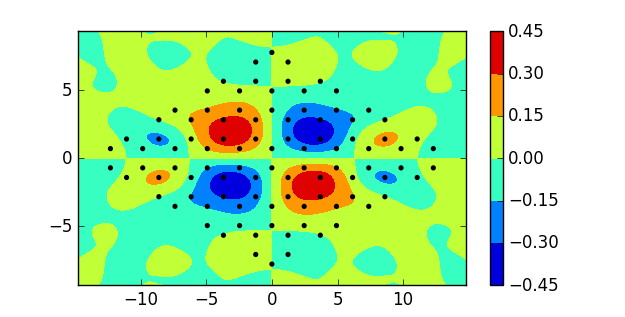}
    \includegraphics[width=0.47\textwidth]{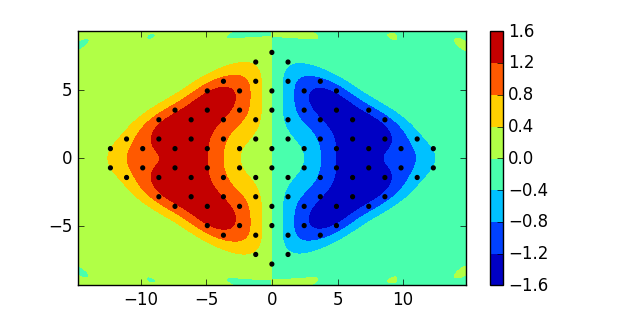}
    \caption{The smooth component of the co-densities $\psi_{46}\psi_{47}$ and $\psi_{46}\psi_{48}$ from Fig.~\ref{fig:codensities}: $2\Re(\Psi_{46,A}^*\Psi_{47,A}+\Psi_{46,B}^*\Psi_{47,B})$ (left) and  $2\Re(\Psi_{46,A}^*\Psi_{48,A}+\Psi_{46,B}^*\Psi_{48,B})$ (right). Note the difference in scale maxima between the two plots.}
    \label{fig:SmoothCodensities}
\end{figure}

Let us focus on the smooth component of each co-density from Fig.~\ref{fig:codensities}, which can be expressed in terms of the envelope functions. For $\psi_{46}\psi_{47}$ and $\psi_{46}\psi_{48}$ they are given by
$2\Re(\Psi_A^*\tilde\Psi_A)\pm2\Re(\Psi_B^*\tilde\Psi_B)$,
and shown in Fig.~\ref{fig:SmoothCodensities}. For $\psi_{46}\psi_{47}$ the smooth component is much weaker, leading to the smallness of the corresponding direct Coulomb integral $V^d_{23}$. The reason for this is the reflection symmetry $\mathcal{R}_{AB}$. Indeed, with respect to this reflection, LUMO and LUMO${}+1$ have opposite parity (similarly to HOMO and HOMO$-1$). Then, $\Psi_B^*(\mathbf{r})\tilde\Psi_B(\mathbf{r})=-\Psi_A^*(\mathcal{R}_{AB}\mathbf{r})\tilde\Psi_A(\mathcal{R}_{AB}\mathbf{r})$, so the smooth part of the co-density $\psi_{46}\psi_{47}$ is $\Re\Psi_A^*(\mathbf{r})\tilde\Psi_A(\mathbf{r})-\Re\Psi_A^*(\mathcal{R}_{AB}\mathbf{r})\tilde\Psi_A(\mathcal{R}_{AB}\mathbf{r})$. Since $\Psi_{A.B}$ and $\tilde\Psi_{A,B}$ correspond to the lowest-energy orbitals, they have no oscillations (see Figs.~\ref{fig:RHOMBenvelopesReIm47} and \ref{fig:RHOMBenvelopesReIm46}, where $\Psi_{47}$ is dominated by the imaginary part and $\Psi_{46}$ by the real part, which have no zeros inside the GQD), so taking the difference between a point $\mathbf{r}$ and its mirror image $\mathcal{R}_{AB}\mathbf{r}$ leads to a strong cancellation. This cancellation also persists when the mirror symmetry is not exact. As a result, the $\phi_H \phi_{H-1}  =  \phi_L \phi_{L+1}$ co-densities are dominated by the fast oscillating $e^{\pm{2}\rmi\mathbf{K}\mathbf{r}}$ components which leads to the smallness of corresponding direct Coulomb integral~$V^d_{23}$.

\end{document}